\documentclass[fleqn,usenatbib]{mnras}

\usepackage{newtxtext,newtxmath}

\usepackage[T1]{fontenc}
\usepackage{ae,aecompl}

\usepackage{graphicx}	
\usepackage{amsmath}	
\usepackage{amssymb}	

\newcommand{\tic}{TIC-231005575}	
\newcommand{\ticc}{TIC-231005576}
\newcommand{\tess}{\textit{TESS}}	%
\newcommand{\ngts}{\textit{NGTS}}	%
\newcommand{\lco}{\textit{LCO}} %
\newcommand{\coralie}{\textit{CORALIE}} %

\title[Photometric recovery of a single-transit candidate]{A long period ($P = 61.8$-d) M5V dwarf eclipsing a Sun-like star from \textit{TESS} and \textit{NGTS}}

\author[0000-0002-4259-0155]{
\parbox{\textwidth}{Samuel Gill$^{1,2}$, 
Benjamin F. Cooke$^{1,2}$,
Daniel Bayliss$^{1,2}$,
Louise~D.~Nielsen$^{3}$, 
Monika~Lendl$^{3,4}$,
Peter J. Wheatley$^{1,2}$,
David R. Anderson$^{1,2}$,
Maximiliano~Moyano$^{13}$,
Edward~M.~Bryant$^{1,2}$,
Jack~S.~Acton$^5$,
Claudia~Belardi$^5$,
Fran\c{c}ois~Bouchy$^{3}$,
Matthew~R.~Burleigh$^5$,
Sarah~L.~Casewell$^5$,
Alexander~Chaushev$^7$, 
Michael~R.~Goad$^5$,
James A. G. Jackman$^{1,2}$,
James~S.~Jenkins$^{11,12}$,
James McCormac$^{1,2}$,
Maximilian~N.~G{\"u}nther,$^{8,9}$
Hugh P. Osborn$^{8,16}$,
Don Pollacco,$^{1,2}$,
Liam~Raynard$^5$,
Alexis~M.~S.~Smith$^{14}$,
Rosanna~H.~Tilbrook$^5$,
Oliver~Turner$^{3}$, 
St\'{e}phane~Udry$^{3}$, 
Jose~I.~Vines$^{11}$,
Christopher A. Watson$^{15}$,
Richard G. West$^{1,2}$,
 }\\
$^{1}$ Department of Physics, University of Warwick, Gibbet Hill Road, Coventry CV4 7AL, UK \\
$^{2}$ Centre for Exoplanets and Habitability, University of Warwick, Gibbet Hill Road, Coventry CV4 7AL, UK \\ 
$^{3}$Observatoire de Gen{\`e}ve, Universit{\'e} de Gen{\`e}ve, 51 Ch. des Maillettes, 1290 Sauverny, Switzerland \\
$^4$ Space Research Institute, Austrian Academy of Sciences, Schmiedlstr. 6, 8042 Graz, Austria\\
$^5$ School of Physics and Astronomy, University of Leicester, Leicester LE1 7RH, UK\\
$^6$ Centre for Exoplanet Science, SUPA, School of Physics and Astronomy, University of St Andrews, St Andrews KY16 9SS, UK\\
$^{7}$Center for Astronomy and Astrophysics, TU Berlin, Hardenbergstr. 36, D-10623 Berlin, Germany\\
$^{8}$Department of Physics, and Kavli Institute for Astrophysics and Space Research, Massachusetts Institute of Technology\\
Cambridge, MA 02139, USA \\
$^{9}$Juan Carlos Torres Fellow\\
$^{10}$ Astrophysics Group, Keele University, Staffordshire, ST5 5BG, UK\\
$^{11}$ Departamento de Astronom\'ia, Universidad de Chile, Camino El Observatorio 1515, Las Condes, Santiago, Chile\\
$^{12}$ Centro de Astrof\'isica y Tecnolog\'ias Afines (CATA), Casilla 36-D, Santiago, Chile\\
$^{13}$Instituto de Astronom\'ia, Universidad Cat\'olica del Norte, Angamos 0610, 1270709, Antofagasta, Chile. \\
$^{14}$Institute of Planetary Research, German Aerospace Center, Rutherfordstrasse 2, 12489 Berlin, Germany\\
$^{15}$Astrophysics Research Centre, School of Mathematics and Physics, Queen's University Belfast, BT7 1NN, Belfast, UK\\
$^{16}$NCCR/Planet-S and Centre for Space and Habitability, University of Bern, Bern 3012, Switzerland\\
}

\date{Accepted XXX. Received YYY; in original form ZZZ}

\pubyear{2019}

\begin{document}
\label{firstpage}
\pagerange{\pageref{firstpage}--\pageref{lastpage}}
\maketitle

\begin{abstract}
The Transiting Exoplanet Survey Satellite (\tess) has produced a large number of single transit event candidates which are being monitored by the Next Generation Transit Survey (\ngts). We observed a second epoch for the \tic\ system (T$_{\rm mag} = 12.06$, T$_{\rm eff} = 5500 \pm 85\,K$) with \ngts\ and a third epoch with Las Cumbres Observatory's (LCO) telescope in South Africa to constrain the orbital period ($P = 61.777\,d$). Subsequent radial velocity measurements with \coralie\ revealed the transiting object has a mass of $M_2 = 0.128 \pm 0.003 M_{\sun}$, indicating the system is a G-M binary. The radius of the secondary is $R_2 = 0.154 \pm 0.008 R_{\sun}$ and is consistent with MESA models of stellar evolution to better than 1-$\sigma$.
\end{abstract}



\section{Introduction}

The Transiting Exoplanet Survey Satellite \citep[\tess,][]{2015JATIS...1a4003R} is well into its primary mission having finished its observations of the southern ecliptic and moved onto the north. However, there are still many discoveries to be found in the first hemisphere of data of which the \tess\ Object of Interest (TOI) catalogue just scrapes the surface. The TOI catalogue is heavily biased towards short period systems that exhibit many transits within their remit of \tess\ data. However, \tess\ data provides an excellent hunting ground for single transit systems \citep{2018A&A...619A.175C,2019AJ....157...84V,2019A&A...631A..83C}. \tess\ single transit systems have, by necessity, periods of greater than $\sim$\,15\,days. Recovering such signals based on a single transit is difficult, though the results are scientifically very interesting. Around M-stars, planets at these periods may be in the temperate zone and longer period eclipsing binaries are of interest as they are less likely to be under the influence of strong tidal interactions. Recently it has been shown that recovery of \tess\ single transits is possible and practical for specialised facilities \citep{2020MNRAS.491.1548G,2019arXiv191005050L}. This is allowing us to begin probing more of these longer period planets and stellar binaries using facilities such as the Next Generation Transit Survey \citep[NGTS,][]{2018MNRAS.475.4476W}.

Continuing to probe systems with larger orbital periods will enable us to learn about the different types of planets, brown dwarfs, and stellar binaries as well as to examine the transition regions between them. \textit{Kepler} was successful in finding planets within their stars temperate zones; the region around a star whereby liquid water could remain stable if an appropriate planetary atmosphere is present \citep{1953ccec.book.....S}. The observing strategy of \tess\ is such that planetary systems identified from a single sector will have orbital periods below 15 days and only reside within the temperate zone if the host is a late M-dwarf. Planets in the temperate zone of more massive stars will have wider orbital separations and longer times between potential eclipses; such systems may transit only once during \tess\ observations. The monotransit Working Group has been established within the Next Generation Transit Survey \citep[\ngts; ][]{2018MNRAS.475.4476W} to recover the orbital period and physical properties of single transit candidates discovered by \tess. The strategy of the working group is to use \ngts\ to monitor \tess\ single-transit candidates with radii below $<1.5\,R_{\rm Jup}$ and recover subsequent epochs in which to determine the physical properties of the transiting system. The transiting companion of some single-transit candidates with radii below $<1.5\,R_{\rm Jup}$ are revealed to be stellar in nature owing to the similarity in size of Jovian-like planets and red-dwarfs. This paper lays out our recovery and characterisation of a \tess\ single-transit candidate, \tic, that is revealed to be an M-dwarf eclipsing a G-type host.

\section{Single-Transit Event Detection}\label{sec:SingleTransitDetection}

\begin{table}
\caption{Photometric colours of \tic.}              
\label{tab:colours}      
\centering   
\begin{tabular}{l c}          
\hline\hline                        
Parameter & value\\
\hline 
Gaia Source ID & 4912474299133826560  \\
RA  &  $01^{\rm h}40^{'}01.35^{"}$\\
Dec  & $-54^{\circ}31' 21.98^{"}$\\

\\
G & $12.855639$ \\
BP & $12.855639$ \\
RP & $14.713296$ \\
pmRA [$\rm mas\, \rm yr^{-1}$] & $52.658 \pm 0.038$ \\
pmDec [$\rm mas\, \rm yr^{-1}$] & $*20.588 \pm 0.039$ \\
Parallax [$\rm mas$] & $2.7886 \pm  0.0267$ \\ \\
\\
TESS [T]  & $12.061 \pm 0.010$ \\
APASS9 [B] & $13.289 \pm 0.010$\\
APASS9 [V] & $12.625 \pm 0.030$\\
APASS9 [g'] & $12.903 \pm 0.038$\\
APASS9 [r'] & $12.442 \pm 0.036$\\
APASS9 [i'] & $12.228 \pm 0.037$\\
2MASS [J  & $11.293 \pm 0.040$] \\
2MASS [H]  & $10.981 \pm 0.040$ \\
2MASS [K$_{s}$]  & $11.037 \pm 0.040$ \\
\hline
\end{tabular}
\end{table}

We conducted a systematic search for single-transit events in lightcurves produced from \tess\ full-frame images \citep{2016SPIE.9913E..3EJ} as described by \citet{2020MNRAS.491.1548G}. The G-type Solar analogue \tic\ was observed with Camera 3 during sectors 2 and 3 (2018 Aug 22 - 2018 Oct 18). We identified a transit event from \tic\ in our search of the \tess\ Sector 3 data, at JD\,2458397.77783. The single transit event has a depth of 22\,mmag and a duration of 7 hours. Excluding the transit feature, the light curve of \tic\ shows a RMS of 1.3\,mmag (over a 1-day timescale), so the transit feature is clearly significant. We inspected individual calibrated \tess\ full-frame images for asteroids and searched for known exoplanets or eclipsing binaries which may be the source of the transit event. We found no reason that the transit event is a false-positive. 

We produced a higher-quality lightcurve using the \textsc{eleanor} pipeline \citep{2019PASP..131i4502F} which we use for the rest of this work. This aperture includes \ticc~($T = 14.9277$) 3.25" away at a position angle of 126.93$^{\circ}$ East of North. The difference in magnitude is 2.867\,mag corresponding to 7.13\% third light in the \tess\ transmission filter. We looked at centroiding information from a 15-pixel cut out of the \tess\ full-frame images around \tic\ to see if there was a minor change to the photocenter during the eclipse event. We find no evidence of changes in the photocenter coincident with the eclipse and so we progressed assuming the transit is on the brighter star.

\begin{figure}
    \centering
    \includegraphics[scale=0.5]{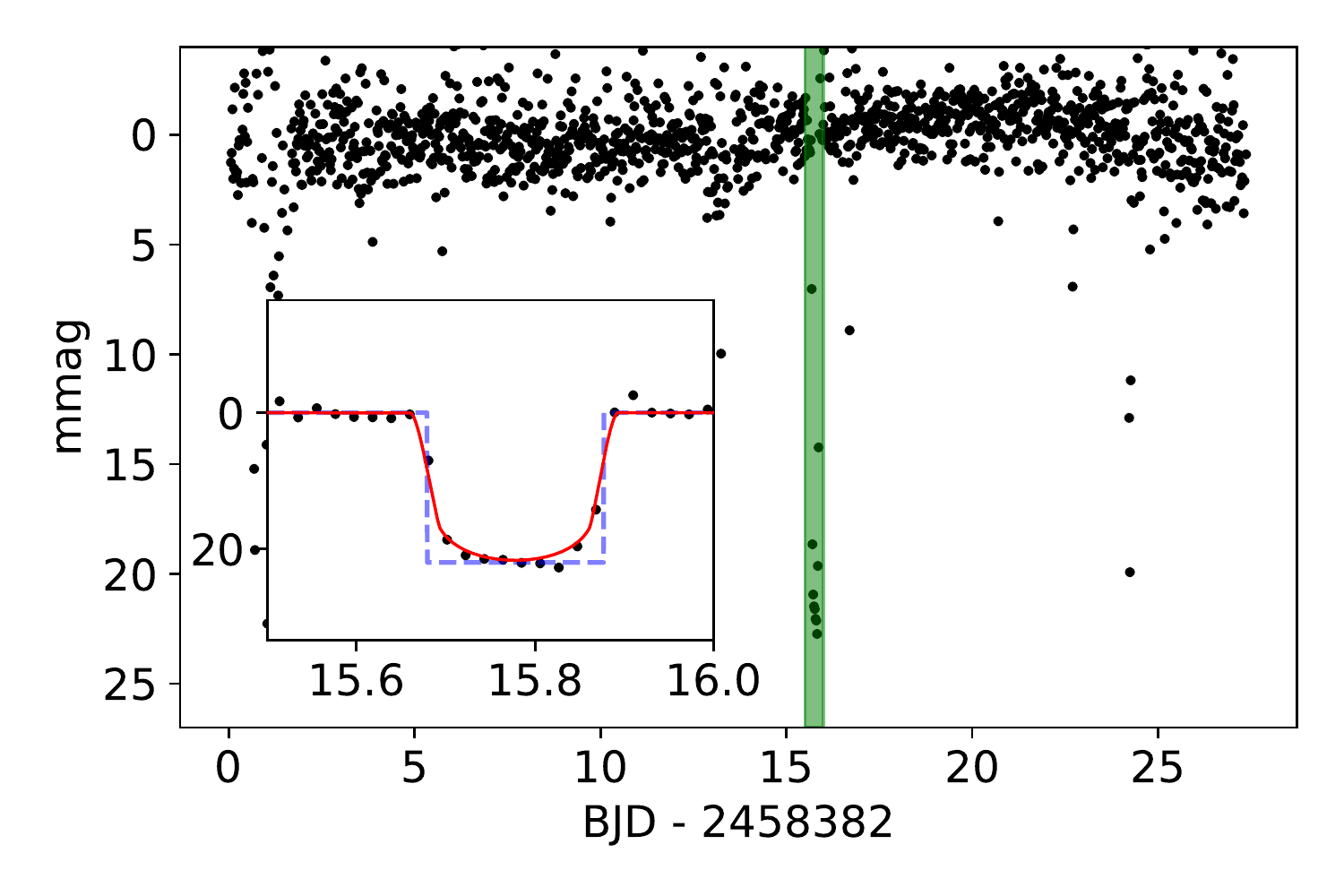}
    \caption{Difference imaging TESS light curve for TIC-231005575 (black). The inset axis shows the transit event highlighted in green, showing the best-fitting global model (red) and box used to detect the single-transit event (blue-dashed). }
    \label{fig:tess}
\end{figure}


\section{A second epoch with NGTS}\label{sec:NGTS_epoch}

We crossmatched \tic\ with archival data from the Wide-Angle Search for Planets \citep[WASP; ][]{2006PASP..118.1407P}. Unlike TIC-238855958 \citep{2020MNRAS.491.1548G}, there are no photometric data points for \tic\ in the WASP archive despite having observations for stars of similar magnitudes within 3 arc minutes of \tic; the reasons for this are unclear. 

In order to recover the orbital period,  we used the \ngts\ telescopes located at the ESO Paranal Observatory in Chile. \ngts\ was designed for very high precision time-series photometry of stars, and thus is the perfect instrument to use for photometric follow-up of \tess\ single-transit candidates. Each NGTS telescope has a field-of-view of 8 square degrees, providing sufficient reference stars for even the brightest \tess\ candidates.  The telescopes have apertures of 20\,cm and observe with a custom filter between 520-890\,nm.  Full details of the \ngts\ telescopes, cameras, and transmission throughput can be found in \citet{2018MNRAS.475.4476W}.

The monotransit working group established within \ngts\ was commissioned to determine the physical properties of systems that appear to transit only once in \tess\ observations. Each target is monitored using a single \ngts\ telescope and is one of at least 12 single-transit candidates observed each night. The working group's strategy is as follows:
\begin{enumerate}
    \item Monitor a \tess\ single transit candidate with \ngts\ until a second transit epoch is detected.
    \item Stop monitoring a target with a second epoch and calculate the predicted epochs for the possible orbital period aliases.
    \item Attempt to observe a third epoch corresponding to possible aliases of the orbital period to confirm the period of the system.
    \item Simultaneously obtain spectroscopic observations for those with a second transit epoch to aid recovery of the orbital period and yield stellar atmospheric properties.  
\end{enumerate}

We started monitoring \tic\ with \ngts\ on the night of 2019 Jul 14.  We observed \tic\ with 10-s exposures when the airmass was below 2 and data were reduced on-site the following day using standard aperture photometry routines. We used the template matching algorithm described in \citet{2020MNRAS.491.1548G} using the transit template to automatically search newly obtained \ngts\ photometric observations for transit events. The transit template was created by modelling the \textsc{eleanor} lightcurve assuming an orbital period of 60 days with limb-darkening parameters interpolated from the effective stellar temperature reported in TESS Input Catalogue 8 \citep{2019AJ....158..138S} assuming solar surface metalicity ([Fe/H]) and surface gravity ($\log g$). The  expected values of $\Delta\log\mathcal{L}$ from transit injection tests allowed for a threshold $\Delta\log\mathcal{L} > 200$.  We observed \tic\ for 25 nights (35,467 exposures) before a second transit event was detected ($\Delta\log\mathcal{L} = 952$) centred at JD=2458706.66152 (see Fig. \ref{fig:Figure_3}).

\begin{figure}
    \centering
    \includegraphics[scale=0.55]{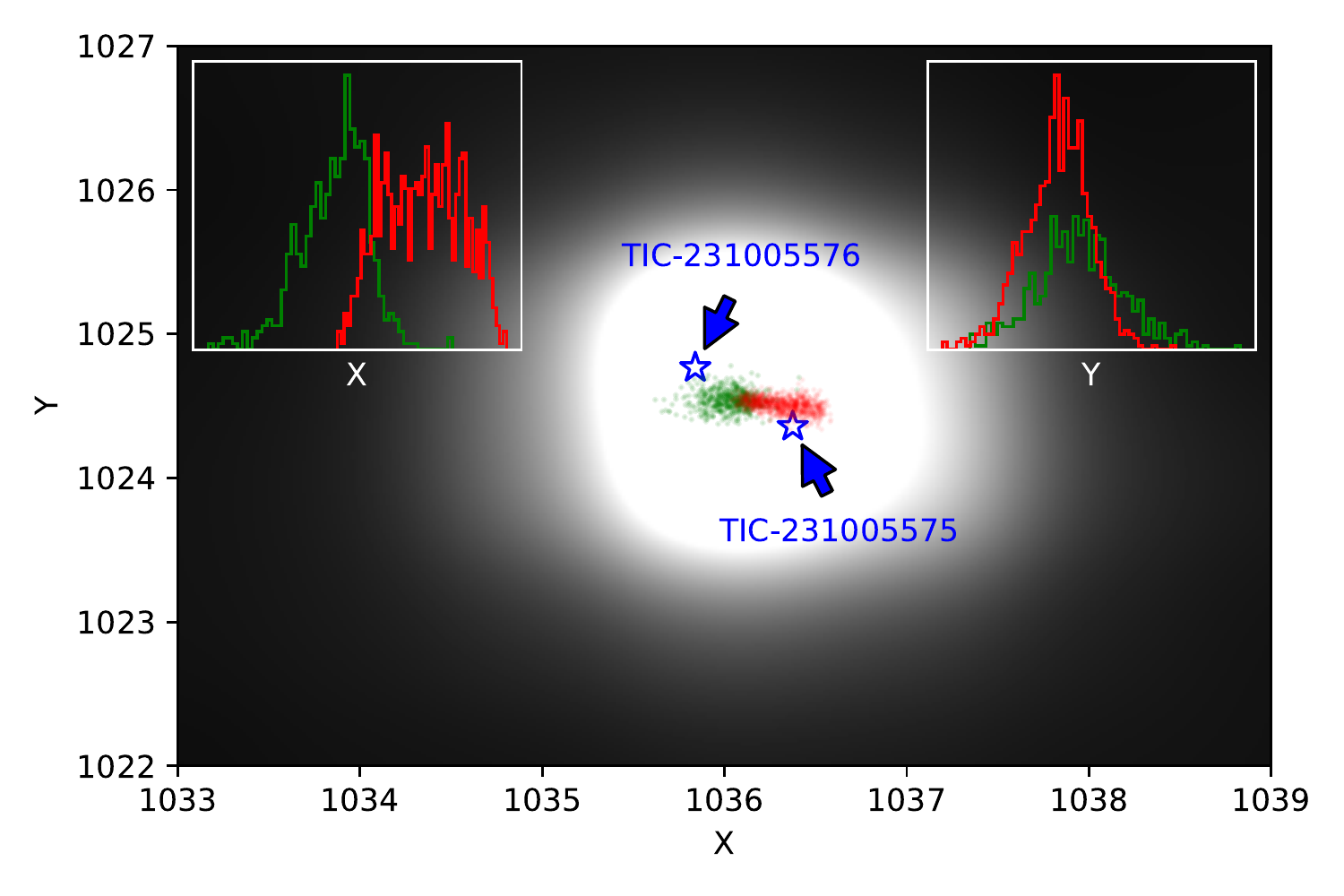}
    \caption{The Gaussian-interpolated \ngts\ reference image with \tic\ and \ticc\ marked (blue stars). For the night of the transit detection (August 11$^{th}$, 2019) we show the in-transit (green) and out-of-transit (red) centroid positions. Histograms of the X and Y centroid positions are shown in their respective subplots. }
    \label{fig:Figure_2}
\end{figure}

The second transit event with \ngts\ contained approximately half the data in-transit and half out-of-transit. The finer plate-scale of \ngts\ combined with sub-pixel centroid positions for \tic\ during aperture photometry provided an opportunity to discern if the transit occurred on \tic\ or \ticc\ (Fig. \ref{fig:Figure_2}). The centroids within transit were closer to \ticc\ and those out out-of-transit were closer to \tic. This indicated that \tic\ was the eclipsing star.

\section{Constraining the orbital period with \lco}\label{sec:LCO_epoch}

The transit epoch from \tess\ and the second recovered epoch from \ngts\ are separated by 308.88353 days. The true orbital period can be no longer than 308 days but can be integer divisions smaller (aliases of the orbital period). Aliases that are permitted depend on the photometric baseline of observations with \tess\ and \ngts. We established that the orbital period could be one of seven orbital periods: 308.88353, 154.44183, 102.96105, 77.22086, 61.77665, 51.48062, and 44.12619 days. Smaller aliases of the orbital period would have been observed in either \tess\ or \ngts\ monitoring observations. 

Establishing the real orbital period required further, time-critical observations of \tic. The first opportunity arose on the night of 2019-09-23 for the 44.13-day alias from Cerro Paranal with \ngts; this did not go ahead due to technical issues. The second opportunity arose on the night of 2019-10-11 for the 61.77-d alias from the South African Astronomical Observatory (SAAO). We scheduled Las Cumbres Observatory 1-m telescope node \citep{2013PASP..125.1031B} at SAAO to observe \tic\ between 19:30 UT and 23:51 UT on the night of 2019-10-11. We obtained 107 science frames using a $r'$ filter with exposure times of 120s and a defocus of 2\,mm. Photometry of \tic\ was extracted using standard aperture photometry routines producing a lightcurve with RMS of 2.17\,mmag (over 30 minutes in-transit)  where a clear partial transit can be seen (see Fig. \ref{fig:Figure_3}). This observation confirmed the 61.77-day alias is the only possible orbital period for \tic.


\section{Spectroscopic Observations}\label{sec:spec}

\begin{table}
\caption{Radial velocity observations of \tic\ from \coralie.}              
\label{tab:radial}      
\centering   
\begin{tabular}{l c}          
\hline\hline                        
JD & Radial velocity [$\rm km\, \rm s^{-1}$]\\
\hline 
2458713.713526 & $-20.1256 \pm 0.0613$ \\
2458717.730527 & $-15.5198 \pm 0.0569$ \\
2458722.798131 & $-13.3451 \pm 0.0408$ \\
2458730.776476 & $-11.8808 \pm 0.0690$ \\
2458737.787439 & $-11.3921 \pm 0.0550$ \\
2458751.680483 & $-11.4952 \pm 0.1370$ \\
2458754.869375 & $-11.8697 \pm 0.1041$ \\
2458776.609257 & $-18.1895 \pm 0.1187$ \\
2458784.625141 & $-13.4274 \pm 0.1545$ \\
2458815.599906 & $-11.5749 \pm 0.0832$ \\
2458839.533646 & $-16.9631 \pm 0.1226$ \\
2458885.523236 & $-12.6690 \pm 0.2317$ \\
2458889.524524 & $-15.7968 \pm 0.1642$ \\
\hline
\end{tabular}
\end{table}

Following the successful recovery of the orbital period of \tic\ using \ngts\ and \lco, we took ten $600$\,s spectroscopic observations of \tic\ using \coralie\ - a fiber-fed \'{e}chelle spectrograph installed on the 1.2-m Leonard Euler telescope at the ESO La Silla Observatory \citep{2001A&A...379..279Q}.  The spectra were reduced using the standard reduction pipeline, and radial velocity measurements derived from standard cross-correlation techniques with a numerical G2 mask. This data is presented in Table \ref{tab:radial} and plotted in Fig. \ref{fig:Figure_3}. We found a semi-amplitude consistent with a stellar transiting companion on an eccentric orbit. We inspected potential dependencies between radial velocities and bisector spans and found little evidence of correlation.


\section{Analysis}\label{sec:analysis}

\subsection{Stellar atmospheric parameters}

\begin{table}
\caption{Stellar atmospheric parameters of the primary G-star, orbital solution, and physical properties of the \tic\ system. Symmetric errors are reported with $\pm$ and asymmetric errors are reported in brackets and correspond to the difference between the median and the 16$^{th}$ (lower value) and 84$^{th}$ (upper value) percentile.}              
\label{tab:parameters}      
\centering   
\begin{tabular}{l c}          
\hline\hline                        
Parameter & value\\
\hline 
Spectroscopy \\
$\rm T_{\rm eff}$ $\rm(K)$
& $5500 \pm 85$ \\

$\log g$ (dex)  
& $4.49 \pm 0.13$ \\

$\xi_{\rm t}\, (\rm km\,s^{-1})$
& $1.17 \pm 1.50$ \\

$v_{\rm mac}\, (\rm km\,s^{-1})$
& $4.67 \pm 1.50$ \\

Vsin$i$ (km\,s$^{-1}$)
& $ \leq 0.5$ \\

$\rm [Fe/H]$ & $-0.44 \pm 0.06$ \\ \\

Orbital solution \\
$\rm T_{\rm 0}$ [JD] &  $2458397.777839_{(688)}^{(730)}$\\
Period [d] & $61.777360_{(163)}^{(179)}$ \\
$R_1 / a$ & $0.0426_{(15)}^{(5)}$\\
$R_2 / R_1$ & $0.4440_{(1)}^{(1)}$ \\
$b$ & $0.573_{(68)}^{(42)}$ \\
$\rm h_{\rm 1, R}$ & $0.7791_{(13)}^{(6)}$ \\
$\rm h_{\rm 2, R}$ & $0.8500_{(1)}^{(1)}$ \\
$\rm h_{\rm 1, r'}$ & $0.7316_{(14)}^{(5)}$  \\
$\rm h_{\rm 2, r'}$ & $0.0.8431_{(1)}^{(1)}$\\
$\sigma_{\rm TESS}$  & $0.00093_{(6)}^{(14)}$ \\
$\sigma_{\rm NGTS}$  & $0.00824_{(4)}^{(32)}$ \\
$\sigma_{\rm LCO}$  & $0.00216_{(22)}^{(10)}$ \\
$K_1$ [km\,s$^{-1}$] & $8.108_{(390)}^{(470)}$ \\
$f_s$ & $0.073_{(13)}^{(10)}$ \\
$f_c$ & $-0.799_{(1)}^{(20)}$ \\
$e$ & $0.298_{(4)}^{(1)}$ \\
$\omega$ [$^\circ$] & $-3.9_{(2.1)}^{(0.9)}$ \\
$V_0$ [km\,s$^{-1}$] & $-14.17_{(3)}^{(27)}$ \\
$J$ [km\,s$^{-1}$] & $0.017{(6)}^{(65)}$ \\ \\

Physical properties \\
$M_1$ [$M_{\odot}$] & $1.045 \pm 0.035$ \\
$R_1$ [$R_{\odot}$] & $0.992 \pm 0.050$ \\
$M_2$ [$M_{\odot}$] & $0.128 \pm 0.003$ \\
$R_2$ [$R_{\odot}$] & $0.154 \pm 0.008$ \\
Age [Gyr] & $3.9 \pm 1.2$ \\

\hline
\end{tabular}
\end{table}

We corrected each CORALIE spectra into the laboratory reference frame before co-adding and re-sampling to produce a spectrum between 450-650\,nm with $2^{17}$ values. We use the wavelet method described in \citet{2018A&A...612A.111G} to extract stellar atmospheric parameters. This method can determine $T_{\rm eff}$ to a precision of $85$\,K, [Fe/H] to a precision of 0.06\,dex and $V \sin i$ to a precision of 1.35\,km\,s$^{-1}$. Values of $\log g$ determined from wavelet analysis are imprecise. To overcome this, we used spectral synthesis (with fixed values of $T_{\rm eff}$, [Fe/H] and $V\sin i$) to model the wings of the magnesium triplets and sodium doublet.  Uncertainties for $\log g$ were calculated by perturbing $\log g$ until the solution was no longer acceptable \citep{2019A&A...626A.119G}. All our derived parameters for \tic\ are set out in full in Table \ref{tab:parameters}.

\subsection{Global modelling}

We modelled all photometric datasets with \coralie\ radial velocities. Initial modelling showed that the transit depths from \ngts\ and \lco\ data sets were consistent to better than 1-$\sigma$ and so we decided to fit a common value of $R_2/R_1$. We used the binary star model described by \citet{2020MNRAS.491.1548G} to calculate models of radial velocity and transit photometry. This model utilises the analytical transit model for the power-2 limb-darkening law presented by \citet{2019A&A...622A..33M}. We fit decorrelated limb-darkening parameters $h_1$ \& $h_2$ (from Eqn. 1 \& 2 of \citealt{2018A&A...616A..39M}) with Gaussian priors centred on values interpolated from Table\,2 of \citet{2018A&A...616A..39M} and widths of 0.003 and 0.046 respectively. The subtle difference between \ngts, \tess, and \lco\ transmission filters are such that we fitted independent values of $h_1$ and $h_2$ for each photometric dataset.

Our model vector included the transit epoch, $T_0$, the orbital period, $P$, $R_1 / a$, $k=R_2/R_1$, $b$, independent values of the photometric zero-point, $zp$, $h_1$ and $h_2$ for each filter, the semi-amplitude, $K_1$, and the systematic radial velocity of the primary star, $\gamma$. Instead of fitting the argument of the periastron ($\omega$) and the eccentricity ($e$), we used  $f_c = \sqrt{e} \cos \omega$ and  $f_s = \sqrt{e} \sin \omega$ since these have a uniform prior probability distribution and are not strongly correlated with each other. We also include a jitter term added in quadrature to radial velocity uncertainties ($J$) to account for spot activity, pulsations, and granulation which can introduce noise in to the radial velocity measurements \citep{2006ApJ...642..505F}. This was added in quadrature to the uncertainties associated with each RV measurement. We fit a similar term for each photometric data set, $\sigma$, which was also added in quadrature to photometric uncertainties. We assume a common third light contribution of 7.13\% in all transmission filters.

\begin{figure}
    \centering
    \includegraphics[scale = 0.6]{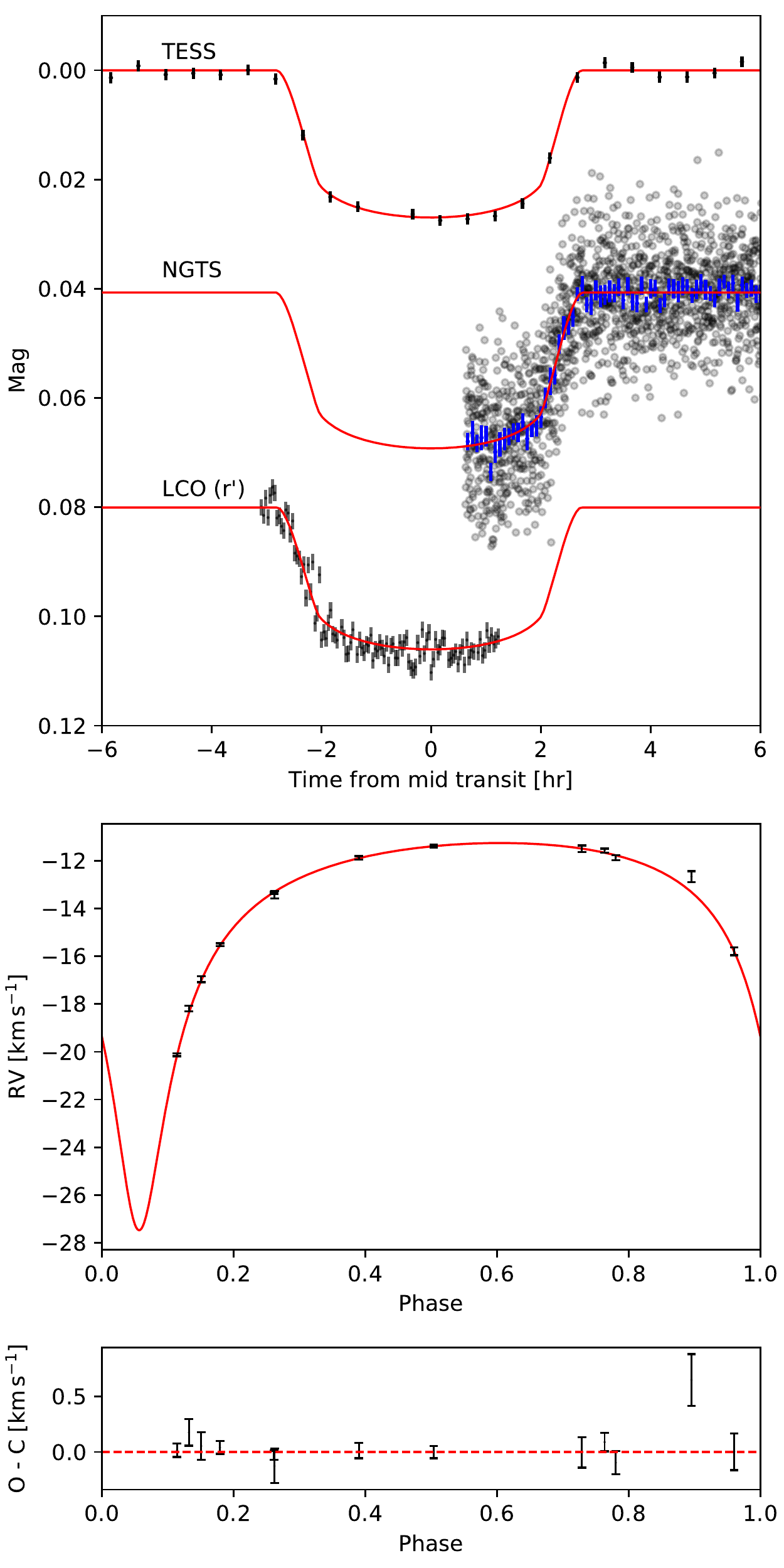}
    \caption{Orbital solution for \tic. Upper panel - Transit photometry (black) for \tess, \ngts, and LCO with best-fitting models (red). For \ngts\ photometry we show the 5-minute binned light curve (blue). Centre panel -- CORALIE radial
    velocity measurements (black) with best-fitting model (red); Lower panel -- fit residuals. }
    \label{fig:Figure_3}
\end{figure}

We used the ensemble Bayesian sampler \textsc{emcee} \citep{2013PASP..125..306F} to sample parameter space. We initiated 50 Markov chains and generated 100,000 trial steps, discarding the first 50,000 steps as part of the burn-in phase. We visually inspected each Markov chain to ensure convergence well before the 50,000$^{th}$ draw. The trial step with the highest log-likelihood was selected as our measurement for each fitted parameter. We adopted the difference between each measured parameter and the $16^{\rm th}$ and $84^{\rm th}$ percentiles of their cumulative posterior probability distributions as a measurement of asymmetric uncertainty. Fitted parameters are reported in Table\,\ref{tab:parameters} and shown in Fig.\,\ref{fig:Figure_3}.

\subsection{Physical properties of \tic}

We used the method described in \cite{2020MNRAS.491.1548G} along with the  \textsc{isochrones} python package \citep{2015ascl.soft03010M} to measure the physical properties of the host star. This method combines Gaia magnitudes $BP$ and $GP$ and parallax with Gaussian priors centred on values reported from GAIA DR2 \citep{2018A&A...616A...1G}, spectroscopically determined values of $\rm T_{\rm eff}$, $\log g$, and [Fe/H], and posterior probability distributions for $e$ and $K_{1}$ to measure the masses, radii, and age of the system.

\section{Discussion}

\begin{figure}
    \centering
    \includegraphics[scale=0.6]{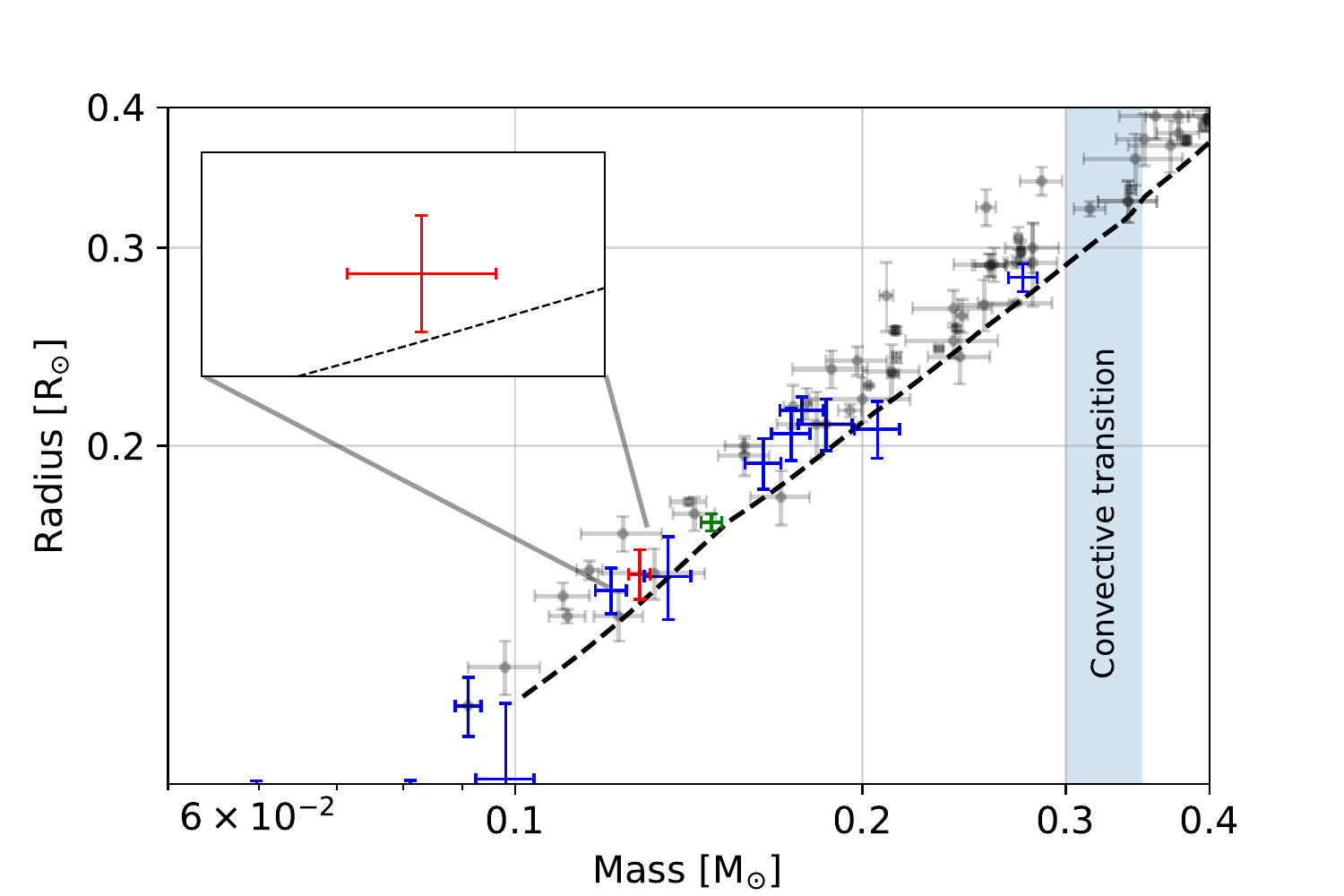}
    \caption{Mass-radius diagram for eclipsing M-dwarfs. The M-dwarf companion of \tic\ is shown in red, with eclipsing M-dwarfs from the EBLM project in blue, and M-dwarfs with mass and radius measurements with less than 10\% uncertainty (from Table 4 of \citet{2018AJ....156...27C}, and references therein) in black.  We also show the mass and radius of TIC-238855958 \citep{2020MNRAS.491.1548G} in green. The best-fitting MESA isochrone for \tic\ (black-dashed) for \tic\ is also shown (black-dashed).}
    \label{fig:my_label}
\end{figure}

\subsection{The \tic\ system}\label{discussion:system}

The primary star in the \tic\ system has a spectral type of G7/8 with physical properties similar to the Sun. Spectral analysis did not reveal anything unusual about the primary star except a relatively metal-poor atmosphere ([Fe/H] = -0.44 $\pm 0.06$) which is approximately 1-$\sigma$ away from the median metalicity of stars from Gaia-ESO data release 3 (\citet{2017ASInC..14...83S}; see Fig. 4 of \citet{2018A&A...612A.111G}). The transiting companion is an M-dwarf with spectral type M5. We interpolated evolutionary models to determine the physical properties of the M-dwarf and found a radius which is inflated by 1.15-$\sigma$ when directly comparing to predicted radius from the best fitting isochrone (0.145\,$R_{\sun}$). A more robust measurement of inflation is discussed in Sect. \ref{discussion:inflation}. The best-fitting radial velocity model resulted in a single radial velocity point (JD = $2458885.523236$) that is $\sim 2$-$\sigma$ higher than expected. The exact reasons for this are unclear, but this point has significantly reduced contrast in the cross-correlation function suggesting moon contamination despite being over 100$^\circ$ away from \tic\ at the time of exposure. Unfortunately, \tic\ has set from Paranal making further spectroscopic observations impossible for this season. 

The proper motion of \tic\ is $\Delta RA = 52.658 \pm 0.038$\,mas\,yr$^{-1}$ and $\Delta Dec =-20.588 \pm 0.039$\,mas\,yr$^{-1}$. \ticc\ is resolved in Gaia (Source ID 4912474299133826688) and has a parallax of $3.0332 \pm 0.0815$ and similar common proper motion of $\Delta RA = 52.699 \pm 0.105$\,mas\,yr$^{-1}$ and $\Delta Dec =-20.592 \pm 0.111 $\,mas\,yr$^{-1}$. \citet{2018A&A...616A...2L} noted that during scanning of close sources the components can become confused due to a changing photocentre. Gaia DR2 assumes that \tic\ and \ticc\ are a single source and they are the primary and secondary source respectively in that solution. We assessed the quality of these astrometric solutions using Eqn.s 1 \& 2 in \cite{2018A&A...616A..17A}. Both solutions pass the first test, but not the second indicating that the astrometric solutions are of poor quality. In addition, astrometric excess noises (\textsc{astrometric\_excess\_noise\_sig}) for \tic\ and \ticc\ are 0\,mas and 30\,mas respectively. This indicates that \tic\ requires no extra noise to the single source solution to fit the observed behaviour, while \ticc\ does. We assume that the astrometric solution for \tic\ is reliable and that the respective solution for \ticc\ is influenced by the proximity and position relative to \tic.


\subsection{Inflation of long-period eclipsing M-dwarfs}\label{discussion:inflation}

\begin{figure}
    \centering
    \includegraphics[scale=0.5]{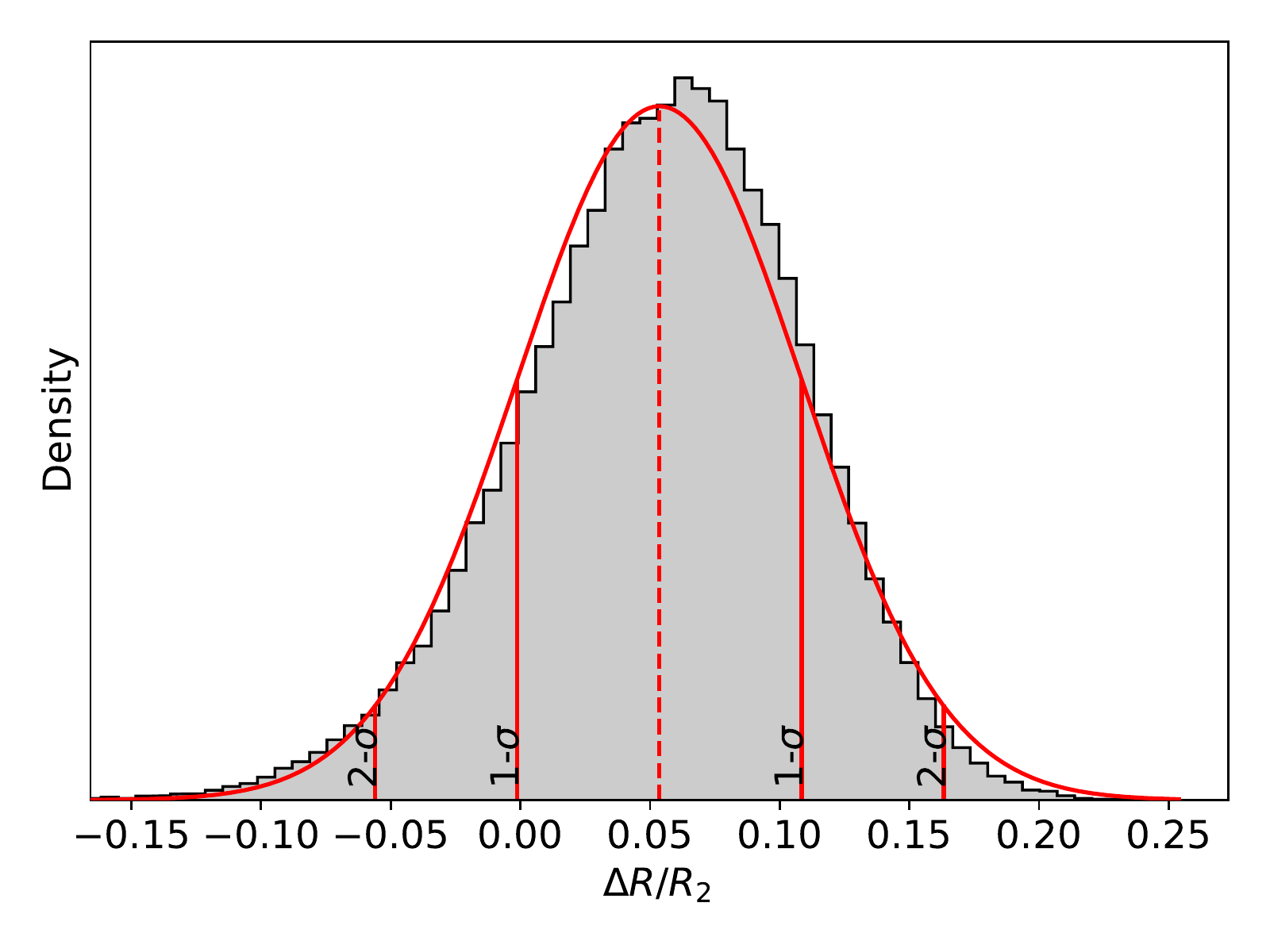}
    \caption{The fractional radius residual PPD for \tic. Red-dashed line marks the measured value of the fractional radius residual and the marked solid red lines indicate the 1-$\sigma$ and 2-$\sigma$ contours.}
    \label{fig:Figure_6}
\end{figure}

There is some tension between measured physical properties of M-dwarfs and predictions from evolutionary models. M-dwarfs across the entire spectral type are reported to have a higher radius than expected by $\sim5$\% \citep{2000ApJ...542..464C,2002ApJ...567.1140T,2003A&A...398..239R,2005ApJ...631.1120L,2008MmSAI..79..562R,2014ApJ...797...31T,2015A&A...577A..42B,2017ApJ...844..134L} and over luminous  \citep{2012MNRAS.423L...1O,2014A&A...572A..50G,2018AJ....156..168B}. This is most apparent for masses whereby M-dwarfs transition from partly-convective to full convective cores \citep[$\sim 0.35$M$_{\odot}$; ][]{2007ApJ...660..732L}.  Magnetic fields are thought to be induced by tidal interactions, enhancing rotation and dynamo mechanisms. This inhibits convection in the core and may be responsible for inflating some stellar radii above those predicted by evolutionary models \citep{2011koa..prop...97K}. However, studies of single M-dwarfs with interferometry \citep{2012ApJ...757..112B} and those in double-lined eclipsing binaries \citep{2012ApJ...757...42F} are comparably inflated by around 3\% making it unclear whether tidal interactions can be blamed \citep{2013ApJ...776...87S}. The \tic\ system is well separated and there is little tidal interaction making it an excellent test of tidally-induced inflation.  

The \tic\ system has a semi-major axis of $23.28 \pm 1.37\,R_{\sun}$. The minimum separation between the primary star and the M-dwarf at perihelion and aphelion is $16.33 \pm 0.96\,R_{\sun}$ and  $30.23 \pm 1.78\,R_{\sun}$ respectively. Consequently, we expect little tidal interaction to occur and so a robust assessment of inflation for this object provides a unique test of models of stellar evolution for an M-dwarf with accurate physical properties in quasi-isolation. Such assessment requires diligent analysis of $M_1$, $R_2$, Age, and [Fe/H] with their respective uncertainties. We follow the method described by \citet{2019A&A...626A.119G} to calculate the posterior probability distribution for the fractional radius residual, $\Delta R_2 / R$, which we briefly describe here. We calculate the posterior probability distribution for the surface gravity of the M-dwarf, $\log g_2$, and combine it with $M_2$ to get a measured value for the radius of the M-dwarf, $R_{2,m}$.  The corresponding draw for age and [Fe/H] was used to interpolate a MESA isochrone \citep{2016ApJS..222....8D,2016ApJ...823..102C} from which an expected radius of the M-dwarf, $R_{2,exp}$, is interpolated when combined with $M_2$. Finally, the posterior probability distribution fractional radius residual compared to MESA isochrones can be calculated,
\begin{equation}
    \frac{\Delta R_2}{R_2} = \frac{R_{2,m} - R_{2,exp}}{R_2}. 
\end{equation}
We calculated the nominal fractional radius residual by binning the posterior probability distribution into 100 bins and fitted a Gaussian model (Fig. \ref{fig:Figure_6}); we took the mean of the fitted Gaussian to be the measurement of $\Delta R_2 / R_2$ with uncertainty equal to the standard deviation. As stated by \citet{2019A&A...626A.119G}, the Gaussian shape is not a perfect fit to the PPDs of $\Delta R_2 / R_2$; there are asymmetric discrepancies where one side of the Gaussian model is lower than the PPD, whilst the other is too high. On average, the under-prediction on one side and over prediction on the other are of the same magnitude and we assume the widths still accurately represent the mean uncertainty of $\Delta R_2 / R_2$. We measured a value of $\Delta R_2 / R_2 = 0.054 \pm 0.055$ and so conclude that the inflation of the  eclipsing M-dwarf in the \tic\ system is not statistically significant (0.98-$\sigma$).

\section{Conclusion}

\tic\ represents the first object to have an orbital period recovered by blind photometric survey as part of the \ngts\ monotransit working Group. \tic\ was initially identified as a single transit candidate from \tess\ differential imaging light curves. The \tess\ single-transit event had shape and depth consistent with a Jovian planet and so was monitored with a single \ngts\ photometer until a second transit event was observed. We excluded all but seven possible aliases of the orbital period which required time-critical photometric observations to either exclude or confirm the true orbital period. We observed a third transit event with \lco\ from Sutherland, South Africa, confirming the 61.77-day orbital period. Spectroscopic observations were used to confirm the primary star's spectral type of G8 with mass and radius consistent with the Sun. 

Joint analysis of photometric and spectroscopic datasets revealed the transiting companion to be a mid M-dwarf ($M_2 = 0.128 \pm 0.003\, M_{\sun}$, $R_2 = 0.154 \pm 0.008\, R_{\sun}$). This is one of the longest period EBLM (eclipsing binary, low mass) systems with accurate physical properties and so we performed a robust assessment of M-dwarf inflation accounting for uncertainties in mass, radius and age of the system. We found that the radius of the eclipsing M-dwarf is consistent with models of stellar evolution to better that 1-$\sigma$.

\section*{Acknowledgements}

The NGTS facility is operated by the consortium institutes with support from the UK Science and Technology Facilities Council (STFC) under projects ST/M001962/1 and ST/S002642/1. 
Contributions at the University of Geneva by FB, LN, ML, OT, and SU were carried out within the framework of the National Centre for Competence in Research "PlanetS" supported by the Swiss National Science Foundation (SNSF).
The contributions at the University of Warwick by PJW, RGW, DLP, DJA, DRA, SG, and TL have been supported by STFC through consolidated grants ST/L000733/1 and ST/P000495/1. DJA acknowledges support from the STFC via an Ernest Rutherford Fellowship (ST/R00384X/1).
The contributions at the University of Leicester by MRG and MRB have been supported by STFC through consolidated grant ST/N000757/1.  SLC acknowledges support from the STFC via an Ernest Rutherford Fellowship (ST/R003726/1)
JSJ is supported by funding from Fondecyt through grant 1161218 and partial support from CATA-Basal (PB06, Conicyt).
MM acknowledges support from the Chilean National Allocation Committee (CNTAC)for the allocation of time on the LCOGT network, semester 2019B proposal id 155475001865. 
JIV acknowledges support of CONICYT-PFCHA/Doctorado Nacional-21191829.
ACC acknowledges support from
the Science and Technology Facilities Council (STFC) consolidated
grant number ST/R000824/1.
MNG acknowledges support from the Juan Carlos Torres Fellowship.
ACh acknowledges the support of the DFG priority program SPP 1992 "Exploring the Diversity of Extrasolar Planets" (RA 714/13-1).





\bibliographystyle{mnras}
\bibliography{references} 







\bsp	
\label{lastpage}
\end{document}